\documentclass[a4paper,english,prl,aps,showpacs,floatfix,groupedaddress,superscriptaddress,twocolumn]{revtex4}
\usepackage[T1]{fontenc}
\usepackage[latin9]{inputenc}
\usepackage{color}
\usepackage{amssymb}
\usepackage{graphicx}

\makeatletter

\pdfpageheight\paperheight
\pdfpagewidth\paperwidth

\@ifundefined{textcolor}{}
{%
 \definecolor{BLACK}{gray}{0}
 \definecolor{WHITE}{gray}{1}
 \definecolor{RED}{rgb}{1,0,0}
 \definecolor{GREEN}{rgb}{0,1,0}
 \definecolor{BLUE}{rgb}{0,0,1}
 \definecolor{CYAN}{cmyk}{1,0,0,0}
 \definecolor{MAGENTA}{cmyk}{0,1,0,0}
 \definecolor{YELLOW}{cmyk}{0,0,1,0}
}

\makeatother

\usepackage{babel}
\begin{document}

\title{Brane parity orders in the insulating state of Hubbard ladders}

\author{Cristian Degli Esposti Boschi}

\affiliation{CNR-IMM, Sezione di Bologna, I-40129, Italy}

\author{Arianna Montorsi}

\affiliation{Institute for condensed matter physics and complex systems, DISAT,
Politecnico di Torino, I-10129, Italy}

\author{Marco Roncaglia}

\affiliation{Institute for condensed matter physics and complex systems, DISAT,
Politecnico di Torino, I-10129, Italy}

\date{\today}
\begin{abstract}
The Mott insulating state of the Hubbard model at half-filling could
be depicted as a spin liquid of singly occupied sites with holon-doublon
quantum fluctuations localized in pairs. In one dimension the behavior
is captured by \textcolor{black}{a finite value of the charge p}arity
string correlator, which fails to remain finite when generalized to
higher dimensions. We recover a definition of parity brane correlator
which may remain nonvanishing in presence of interchain coupling,
by assigning an appropriate fractional phase to the parity breaking
fluctuations. In case of Hubbard ladders at half-filling, we find
that the charge parity brane is non-zero at any repulsive value of
interaction. The spin parity brane instead becomes nonvanishing in
the even-leg case, in correspondence to the onset of the spin gapped
D-Mott phase, which is absent in the odd-leg case. The behavior of
the parity correlators is also analyzed by means of a numerical DMRG
analysis of the one- and two-leg ladder.
\end{abstract}

\pacs{71.10.Fd, 71.10.Pm, 05.30.Rt}

\maketitle

\section{Introduction}

The presence of nonlocal orders (NLO) in various low temperature phases
of quantum matter, ultimately relying on entanglement of short/long-range
degrees of freedom, is a fascinating theoretical prediction \cite{NiRo}
which has recently also been observed in experimental settings \cite{Enal,Paal}.
NLO amount to the nonvanishing expectation value of correlators between
nonlocal operators. The fact seems to contradict Landau's paradigm,
which associates the formation of ordered phases to the breaking in
the ground state of some symmetries of the Hamiltonian, and to the
corresponding finite asymptotic value of appropriate correlators between
local observables identified as order parameters. NLO connect with
the presence of topological orders, which have been thoroughly investigated
in the last two decades since their identification in edge states
of fractional quantum Hall liquids \cite{Wen}, to the recent classification
of topological phases of noninteracting fermions \cite{Tis}. In case
of quantum chains of interacting spins, general results have been
achieved for one dimensional systems \cite{Ptbo,Cgw}, introducing
the concept of symmetry protected topological (SPT) phases. These
have been subsequently extended to interacting fermionic systems \cite{Tpb}
(see also \cite{Wen-1} for a review).

Nonlocal correlators appear to play the role of order parameters for
SPT phases \cite{Potu,DuQu}. However, phases with NLO may be trivial
from the point of view of topological order: this is the case for
instance for the Mott insulator (MI) in the one-dimensional (1D) Hubbard
model, characterized by a nonvanishing value of the charge parity
correlator in the asysmptotic limit, both in the bosonic \cite{BDGA}
and in the fermionic case \cite{Moro}. The 1D parity correlator is
the string product of site parities connecting two arbitrary points
in the lattice. More generally, a bosonization analysis has proved
that NLO configure as order parameters for the ordered phases of spinful
fermions in 1D with time reversal symmetry: each pinned value of either
the spin or charge bosonic fields corresponds to the presence of specific
nonvanishing NLO \cite{Bmr}. A fact that suggests a possible fundamental
role of NLO in fermionic systems. Also, since string correlators are
products of quantities on physical sites, they can be easily measured
in experimental setups on optical lattices, representing a feasible
way to distinguish phases with and without topological order \cite{BaVi}.

The generalization of the above concepts to higher dimension is not
straightforward. It has been argued that string orders and SPT phases
are fragile with respect to interchain tunneling \cite{AnRo2,AnRo,Mopo},
when the system breaks reflection symmetry. Even in presence of such
symmetry, in \cite{AnRo} it was shown that string-like correlations
decay to zero with an area law in the two-dimensional (2D) case: the
proof holds for correlators between points. A different definition
of the parity correlator in 2D \cite{RSEZ} relies instead on a generalization
of string to branes when moving to 2D systems. In this case the parity
correlator was shown to exhibit a\textit{ }``perimeter law'' decay
to zero within the MI phase. 

While some specific definition of nonlocal order may be evanescent
in higher dimension, the underlying physics could persist in appropriate
phases \cite{BDGA,End}. In case of parity, it is related to the presence
of finite-size localized pairs of parity-changing objects which as
a whole do not destroy the parity of the background. The characterization
of such behavior by means of observables, as well as its identification
in appropriate physical systems, is a relevant open issue.

\section{Fractional parity branes}

Here we revisit the definition of NLO, by specializing to parity orders,
with the scope of capturing their role in the onset of insulating
behavior in higher dimension. We focus onto the 2D fermionic case,
though most of the concepts introduced can be easily generalized to
higher dimension. According to the analysis developed in \cite{Moro},
at least two types of parity correlators can be identified in these
systems, thanks to the conservation of both charge and spin. We adapt
to spin and charge parity correlators a definition first introduced
in \cite{Nhs} for detecting Haldane-type orders. This amounts to
considering in 2D branes of operators bounded by rungs rather than
1D strings bounded by sites; at the same time assigning a phase term
$\exp(i\theta)$ to the parity-destroying elements. We show that increasing
the number of legs, $\theta$ should be scaled down with that number,
in order to capture the presence of parity order in the corresponding
gapped phases. We then adopt the definition of such nonlocal fractional
parity brane correlators to the study at half-filling of the repulsive
Hubbard model on a chain and a two-leg ladder.

\subsection{\textit{Definition }}

We think of a $M{\rm x}N$ lattice as a $N$-rung, $M$-leg ladder.
The lattice is folded onto a torus by assuming periodic boundary conditions
(PBC) in both directions. The total Hilbert space is the direct product
${\cal H=}\bigotimes_{k}{\cal H}_{k}$ of local Hilbert spaces ${\cal H}_{k}$
defined on the rungs ($k=1,\dots,N$). In general, each symmetry of
the model Hamiltonian can be decomposed into irreducible representations
with well defined Cartan generators. Let us focus on a specific local
Cartan generator $h_{k}$ with highest weight $\Lambda$. We define
the parity string operators
\begin{equation}
O_{P}^{(h)}(j)=\prod_{k<j}\exp\left(i\pi\frac{h_{k}}{\Lambda}\right).\label{eq:str_op}
\end{equation}
The local operator $h_{l}$, assumes $2\Lambda+1$ discrete values
in the interval $-\Lambda\leq h_{l}\leq\Lambda$. We now introduce
the string correlators
\begin{equation}
{\cal C}_{P}^{(h)}(r)=\left\langle O^{(h)\dagger}(j)O^{(h)}(j+r)\right\rangle ,\label{eq:2dcorr}
\end{equation}
and we say that we are in presence of parity hidden order when ${\cal C_{P}}^{(h)}(r)$
is nonvanishing in the limit $r\rightarrow\infty$. It is important
to notice that in our definition (\ref{eq:str_op}) the eigenvalues
of the Cartan generators in the local representation have been renormalized
by $\Lambda$, so that the phase factor in the exponent varies in
the interval $[-\pi,\pi]$. This choice is crucial to avoid unwanted
phase cancellations along the rungs that would came up without the
division by $\Lambda$ in Eq.(\ref{eq:str_op}), especially when the
rung size $M$ is increased and accordingly also $\Lambda$ grows
(depending on the irrep we are considering). Our choice is at odd
with a majority of the string operators studied in the recent literature
(see for instance \cite{RSEZ}) though consistent with some previous
literature on spin systems, like Ref. \cite{Nhs} where a Haldane-string
of spins ($h_{k}=S_{k}^{z}$, in that case) was studied in a two-leg
ladder. As an example which we will thoroughly discuss, in the fermionic
Hubbard chain with charge and spin conservation ${\cal C}_{P}^{(h)}(r)$
coincides with the so called parity correlators \cite{Moro,Enal},
with $k$ single site index, $h_{k}\equiv S_{\nu k}^{z}$, $\nu=c,s$,
and $S_{ck}^{z}=(n_{k\uparrow}+n_{k\downarrow}-1)/2$, $S_{sk}^{z}=(n_{k\uparrow}-n_{k\downarrow})/2$
the pseudo-spin and spin operator respectively; and, of course, $\Lambda=1/2$. 

The definition (\ref{eq:str_op}) is decisive for the persistence
of string order when we consider the 2D limit $M\rightarrow\infty$
\begin{equation}
{\cal C}_{P}(h)\doteq\lim_{M\rightarrow\infty}\lim_{r\rightarrow\infty}\,{\cal C}_{P}^{(h_{M})}(r)\label{eq:2dcorr tl}
\end{equation}
where $h_{M}$ is a Cartan generator of the symmetry algebra of the
Hamiltonian model for a rung of length $M$. In the following section
we give some arguments to support our choice.

\subsection{Diluted empty-doublon pairs limit}

Close to the atomic limit, approached for large interactions, we can
consider empty-doublon fluctuations as dilute and independent. Let
us consider a domain limited by two rungs of total length $2M$. Paired
fluctuations that lie entirely inside or outside such a domain, do
not contribute to a phase term in Eq.(\ref{eq:str_op}). The terms
that are candidate to destroy the parity order are given by pairs
where one partner lies inside the domain, while the other is outside.
In the standard parity order, every ``cut'' gives a $-1$, so the
parity order is given by the difference of all the configurations
with an even number of cuts minus all the configurations with an odd
number. A probability calculation \cite{End} that assumes all the
configutations equally probable (dilution limit), shows that the parity
order vanishes with $M$, with a ``perimeter law''. 

On the contrary, we show that the parity correlator defined in Eq.(\ref{eq:str_op})
remains finite in the gapped phase. In fact, let's confine our analysis
to a symmetry algebra $su(2\nu)$ ($\nu$ positive integer): when
we consider the rung pseudo-spin $S_{\nu}^{z}(j)=\sum_{l=1}^{M}S_{\nu}^{z}(j,l)$
($l$ site- and $j$ rung-index), we recognize that in this case the
highest weight is $\Lambda=M/2$. The fractional parity operators
(\ref{eq:str_op}) become 
\begin{equation}
O_{P}^{(\nu)}(j)=\prod_{k<j}\exp\left[i\frac{2\pi}{M}S_{\nu}^{z}(j)\right].\label{eq:NLOnu}
\end{equation}
With a similar probability argument as the one proposed in \cite{End},
we assign a probability $p$ to find a broken pair in correspondence
of each lattice site along the boundary. In particular, if the partner
inside the domain is an empty site, the contribution is $\exp\left(-i\frac{\pi}{M}\right)$;
if instead the doublon is inside the contribution is $\exp\left(i\frac{\pi}{M}\right)$.
A probability calculation gives a two-point (-rung) correlator (\ref{eq:2dcorr})
\begin{equation}
{\cal C}_{P}^{(\nu)}(r)=\left[1-p\left(1-\cos\frac{\pi}{M}\right)\right]^{2M}.\label{eq:frac parity 2D}
\end{equation}
where the separation $r$ is larger than the typical size of the desordering
pairs. Interestingly Eq.(\ref{eq:frac parity 2D}) tends to 1 in the
limit $M\to\infty$. This promotes the 2D fractional parity correlator
(\ref{eq:2dcorr tl}) as a good order parameter to remain finite in
the MI phase.

In case the rungs are uncorrelated, the same reasoning of the previous
paragraph shows that ${\cal C}_{P}(r)=\left[1-p\left(1-\cos\frac{\pi}{M}\right)\right]^{MN},$
which, upon considering the limit in the order as in equation (\ref{eq:2dcorr tl}),
correctly reproduces a vanishing value for the 2D brane correlator.
In the present formulation, the qualitative result is independent
of the specific 2D domain chosen: we do not necessarily deal with
regions bounded by two rung, but we can depict any collection of closed
domains with perimeter $2M$. The essential fact in order to possibly
get a finite fractional parity is that the ``cut'' disordering pairs
must be localized along the perimeter.

\section{the hubbard model at half filling on a ladder}

The model Hamiltonian reads

\[
H_{Hub}=-\sum_{<i,j>\sigma}c_{i\sigma}^{\dagger}c_{j\sigma}+U\sum_{i}(n_{i\uparrow}-\frac{1}{2})(n_{j\downarrow}-\frac{1}{2}).
\]
At half-filling the symmetry algebra of the Hamiltonian, induced by
total spin and pseudo-spin conservation, is $su(2)\bigoplus su(2)$.
The symmetry algebra has charge and spin Cartan operators, namely
$S_{\nu}^{z}=\sum_{j}S_{\nu j}^{z}$, with $\nu=c,s$ respectively. 

For an infinite value of particle interaction $U$, the ground state
of the fermionic (and bosonic) Hubbard model displays a MI phase in
the repulsive case, where each site contains only one particle, with
parity order equal to unity in the charge sector. For finite positive
values of the interaction, in 1D the MI phase is also characterized
by holon-doublon quantum fluctuations, which remain confined in the
background of singly occupied sites. In the attractive interaction
case a specular description holds by interchanging holon-doublon and
up-down spin single particles roles. The system, at zero magnetization
and arbitrary filling, is described by a Luther Emery liquid (LE):
each site contains either holons or doublons, with quantum fluctuations
of single electrons with up and down spins. The parity order in the
charge sector (which assignes the minus sign to both holons and doublons)
remains asymptotically finite in the MI phase. Whereas in the spin
sector it is vanishing in the MI phase and in the fermionic case the
spin parity order remains finite instead in the LE phase \cite{Moro}. 

In 2D, the systems in the repulsive case is expected to be an antiferromagnetic
insulator with vanishing spin gap. The ``paired fluctuation'' physical
picture of the insulator could hold as well in dimension greater than
one \cite{End}. However, up to now the proposed 2D generalizations
of the parity correlator have failed to remain finite in the corresponding
phases. 

In this section, we apply the definition of spin and charge parity
orders introduced in the previous section to investigate, both by
bosonization and numerical analysis, their behavior in the repulsive
regime of the Hubbard model on ladders.

\subsection{Bosonization analysis of repulsive M-leg ladder }

The bosonization analysis of the Hubbard model on a ladder at half-filling
was first considered in the weak coupling limit in the two-leg case
\cite{Schu,BaFi}, and extended to the generalized Hubbard model case
in \cite{Frad,Furu}. At variance with the single chain case, for
repulsive values of the interaction the insulating phase is fully
gapped: also the spin liquid becomes insulating, and the phase is
denoted as D-Mott. The bosonization analysis was successively extended
to the $M$-leg case in \cite{BaFi2} and \cite{Lede}. The analysis
in the weakly interacting repulsive case shows an even-odd effect.
At appropriate energy scales, for $M$ even the Hamiltonian becomes
the sum of $M/2$ decoupled two-leg ladder Hamiltonians, whereas for
$M$ odd it is the sum of $(M-1)/2$ two-leg ladder plus a single
chain Hamiltonians. Thus the behavior of the $M$-leg ladder can be
derived from that of the single- and two-chain cases. In particular,
below we will refer to the total charge and spin fields on the two-leg
ladder as $\phi_{\nu}^{+}$, and on the single chain as $\sqrt{2}\phi_{\nu}$:
according to the literature \cite{Giam}\cite{Ners}, in the charge
sector they both are locked to the value $0$ (modulo integer multiples
of $\pi$); whereas in the spin sector $\phi_{s}^{+}$ is locked to
the value zero, while $\sqrt{2}\phi_{s}$ is unpinned. So that the
total charge field is always pinned to $0$; whereas the total spin
field in the even $M$ case is pinned to 0, giving rise to a D-Mott
phase, but in the odd $M$ case it is unpinned, recovering the 1D-like
spin liquid MI. Depending on the strength of interaction, a dimensional
crossover is also expected with increasing the number of legs, signaled
by the suppression of the spin gap also in the even leg case, and
to the collapse to the antiferromagnetic insulating phase characteristic
of the 2D system.

We now proceed to investigate how the above features reflect into
the values of the charge and the spin brane parity correlators introduced
in the previous section. Within the bosonization approach, the parity
correlators can be evaluated by noticing that the product of the exponentials
in Eq.(\ref{eq:str_op}) in the continuum limit becomes the exponential
of the integral along the rung-chain of $2\pi S_{\nu}^{z}(x)$, with
the total rung densities $S_{\nu}^{z}(x)\doteq\frac{1}{M}\sum_{l}S_{\nu}^{z}(x,l)$
($\nu=c,s$ respectively). Due to the paired chain structure of the
bosonized problem, we can assume $S_{\nu}^{z}(x)\approx\frac{1}{2}([S_{\nu}^{z}(x,1)+S_{\nu}^{z}(x,2)]\doteq\frac{1}{2\pi}\partial_{x}\phi_{\nu}^{+}$
for even $M$; and, analogously $S_{\nu}^{z}(x)\approx\frac{1}{2\pi M}[(M-1)\partial_{x}\phi_{\nu}^{+}+\sqrt{2}\partial_{x}\phi_{\nu}]$,
for odd $M.$ So that, exploting the particle-hole symmetry, one has

\[
O_{P}^{\nu}(x)\approx\cos\tilde{\phi}{}_{\nu}(x)\quad,
\]
with $\tilde{\phi}{}_{\nu}\equiv\phi_{\nu}^{+}$ for $M$ even, and
$\tilde{\phi}{}_{\nu}\equiv\phi_{\nu}^{+}+\frac{1}{M}(\phi_{\nu}-\phi_{\nu}^{+})$
for $M$ odd. The asymptotic parity correlators defined in (\ref{eq:2dcorr})
can now be evaluated according to the analysis outlined in \cite{Giam}
and \cite{Ners} (see also \cite{Moro}). In full analogy with the
one-chain leg, one obtains:
\[
{\cal C}_{P}^{(\nu)}(r)=\langle\cos\tilde{\phi}_{\nu}(0)\cos\tilde{\phi}_{\nu}(r)\rangle
\]

When both total charge and spin bosonic fields are locked to the value
0, as in the even-leg ladder case, the corresponding brane correlators
${\cal C}_{P}^{(\nu)}$ defined in (\ref{eq:2dcorr tl}) become nonvanishing,
signaling the simultaneous presence of two types of hidden orders:
localized holon-doublon pairs, and localized up-down spin pairs. Whereas,
when a single one-chain field is unlocked, also the associated total
field becomes unpinned, and the corresponding correlator will have
vanishing average. This is the case for the spin field in the odd-leg
ladder case.

\subsection{Numerical analysis of Hubbard chain and two-leg ladder }

\begin{figure}
\includegraphics[width=8cm]{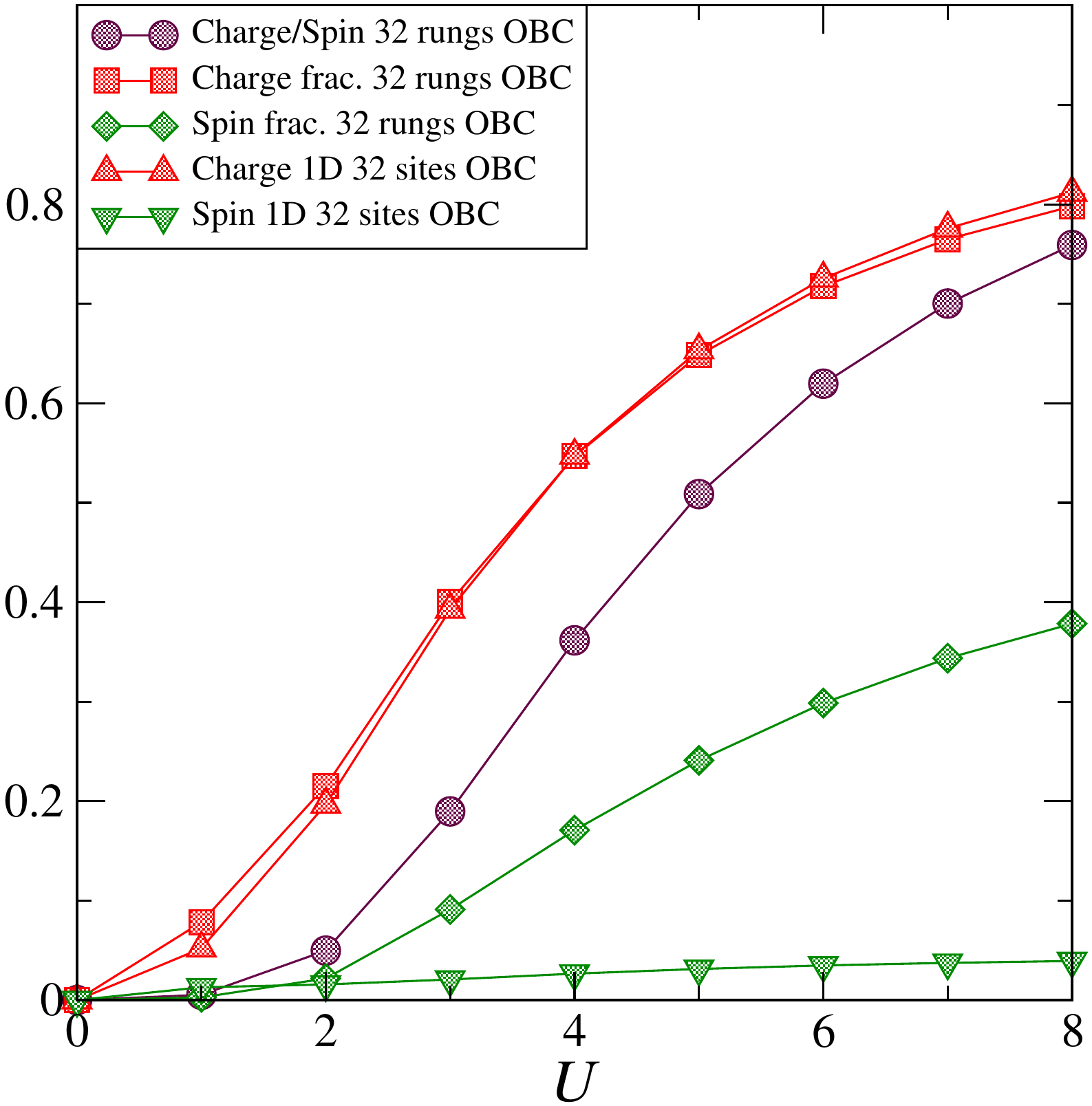}

\caption{Charge (red) and spin (green) fractional (squares) parity brane correlators
for the two-leg ladder with OBC as functions of the repulsive interaction
$U$, compared with the non-fractional definition (maroon circles)\cite{RSEZ},
and the one-leg ladder case (triangles). In the fractional case, increasing
the number of legs, no reduction is observed for the charge parity
correlator. The number of rungs (32) is expected to give a faithful
picture of the thermodynamic limit for $U>2$; finite-size effects
are still present for smaller values. The plotted curves have been
shifted according to the corresponding data at $U=0$. From the arguments
of Fig. 7 of the Supplementary Information\cite{SI} we may consider
to have a vanishing asymptotic value for $U=0$ and a small finite
value for $U=1$ \label{fig:parity-OBC_vs_U-1}.}
\end{figure}

We now proceed to the evaluation of the brane parity spin and charge
correlators, ${\cal C}_{P}^{(\nu)}(r)$ for $\nu=s,c$ and $M=1,2$
with a series of density-matrix renormalization group (DMRG) calculations\cite{SI}.
Note that as long as one uses a ``nonfractional'' expression as
in Ref.\cite{RSEZ} for the single-site single-species operators we
have $\exp({\rm i}\pi n_{i\sigma})=\exp(-{\rm i}\pi n_{i\sigma})$
so we will report only the charge case because the spin case is simply
related to the former. At variance with the case $M=1$ the spatial
dependence of the nonlocal parity correlation functions is not simply
given by a decay with the distance and even-odd oscillations, but
depends in a nontrivial fashion on the number of rungs and on the
actual value of the interaction. After a more detailed discussion
found in the Supplementary Information here we give conventionally
the values computed with open boundary conditions (OBC) from 1/4 to
3/4 of the total ladder length in terms of rungs (furtherly averaged
between the last two rungs to extract the uniform part); the choice
is meant to reduce edge effects and to keep a sufficiently large distance
between the rungs in the correlator. The resulting dependence on $U$
is plotted in Fig.\ref{fig:parity-OBC_vs_U-1} for a total number
of 32 rungs. Within the limits due to numerical errors and to the
finite size, the figure clearly shows that the fractional definition
of the parity brane correlators at the same time mantain constant
its value in the charge sector, and passes from 0 to finite in the
spin sector. Whereas the nonfractional definition would exhibit a
damping of the charge correlator, and an identical behavior of the
spin correlator on the 2-leg ladder.

\section{Conclusions}

We have considered the problem of characterizing the presence and
type of long range order in the Mott insulating phase of Hubbard-like
models in dimension greater than 1. Our results confirm that such
phase can be depicted as a spin liquid of single electrons with localized
correlated quantum fluctuations of holons and doblons. For fermionic
Hubbard models in one dimension, this behavior was already known to
be captured by a nonvanishing value in the asymptotic limit of the
charge string parity correlator between points \cite{Moro,Bmr}. In
the bosonic two dimensional case, a recent theoretical investigation
did show \cite{RSEZ} that a parity brane correlator can be introduced,
which exhibit a perimeter law decay to zero within the charge gapped
phase, at variance with the area law characteristic of the gapless
systems \cite{SwSe}. Here we proposed the fractional definition given
in Eqs. (\ref{eq:str_op})-(\ref{eq:2dcorr tl}) of the parity brane
correlator on ladders, which amounts to normalize the parity to the
actual length of the perimeter. We provided analytical evidence that
such quantity could remain finite in 2D in the asymptotic limit within
the insulating phase, both by investigating the strong coupling dilute
limit, and through the analysis of bosonization results on ladders.
We also proceeded to the numerical investigation of the single chain
and two-leg ladder cases. Already in these simple cases, our definition
of the charge parity correlator does not show appreciable reduction
of its value with increasing the number of legs, at variance with
previous ones. Moreover, by appliying the same type of analysis to
the fractional spin parity brane, it is seen that the opening of the
spin gap in the even legs case is signaled by its finite value, in
contrast with the one-leg ladder. Noticeably, the present definition
allow to observe how charge and spin degrees of freedom could behave
independently up to the 2D limit.

The results also provide a quantitative tool for detecting the occurrence
of the above behavior both in experimental setups \cite{GPMC} and
in numerical simulations of 2D systems of interacting electrons/spinful
fermionic atoms. In fact, within the same framework one could also
study for different parameters' regimes the possible permanence of
the Luther Emery superconducting phase: by analogy with 1D findings\cite{Moro},
the phase may be described by a finite value of the spin parity brane,
accompanied by a vanishing value of the charge brane.

Finally the present picture of the insulating phase of the Hubbard
model can be re-interpreted in terms of entanglement and SPT phases
\cite{Wen-1}. Indeed the phase can be thought of as the infinite-U
insulating phase of singly occupied sites, in which short-range entangled
pairs of holons-doublons are created at finite U by appropriate local
unitary transformations. This suggests a role of brane correlators
as order parameters for distict trivial SPT phases\cite{FPO} in higher
dimension. 
\begin{acknowledgments}
AM acknowledges the kind hospitality of the Condensed Matter Theory
Visitor\textquoteright s Program at Boston University; as well as
useful comments by Senthil Todadri. CDEB acknowledges the use of computing
facilities at INFN Bologna for a part of the DMRG calculations.\end{acknowledgments}

\bibliographystyle{apsrev}
\bibliography{entanglement}

\begin{thebibliography}{10}
\bibitem{NiRo}M. den Nijs, K. Rommelse, Phys. Rev. B \textbf{40},
4709 (1989).

\bibitem{Enal}M. Endres et al., Science \textbf{334}, 200 (2011). 

\bibitem{Paal}J.A.M. Paddison et al., Science \textbf{350}, 179 (2015).

\bibitem{Wen}X.-G. Wen, Adv. Phys. \textbf{44}, 405 (1995).

\bibitem{Tis}A.P. Schnyder, S.R., A. Furusaki, and A.W. Ludwig, Phys.
Rev. B \textbf{78}, 195125 (2008); X.L. Qi, T.L. Hughes, and S.C.
Zhang, \emph{ibid.} 195424 (2008).

\bibitem{Ptbo}F. Pollmann, A.M. Turner, E. Berg, and M. Oshikawa,
Phys. Rev. B \textbf{81}, 064439 (2010).

\bibitem{Cgw}X. Chen, Z.-C. Gu, and X.-G. Wen, Phys. Rev. B \textbf{83},
035107 (2011); \textbf{84}, 235128 (2011).

\bibitem{Tpb}A.M. Turner, F. Pollmann, and E. Berg, Phys. Rev. B
\textbf{83}, 075102 (2011).

\bibitem{Wen-1} B. Zeng, X. Chen, D.-L. Zhou, X.-G. Wen, ``Quantum
Information Meets Quantum Matter -- From Quantum Entanglement to Topological
Phase in Many-Body Systems'', preprint arXiv:1508.02595

\bibitem{Potu}F. Pollmann, and A. Turner, Phys. Rev. B \textbf{86},
125441 (2012).

\bibitem{DuQu}K. Duivenvoorden, and T. Quella, Phys. Rev. B \textbf{86},
235142 (2012).

\bibitem{BDGA} E. Berg, E.G. Dalla Torre, G. Giamarchi, and E. Altman,
Phys. Rev. B \textbf{77}, 245119 (2008).

\bibitem{Moro}A. Montorsi, and M. Roncaglia, Phys. Rev. Lett. \textbf{109},
236404 (2012).

\bibitem{Bmr}L. Barbiero, A. Montorsi, and M. Roncaglia, Phys. Rev.
B \textbf{88}, 035109 (2013).

\bibitem{BaVi} Y. Bahri, and A. Vishwanath, Phys. Rev. B \textbf{89},
155135 (2014).

\bibitem{AnRo2}F. Anfuso, and A. Rosch, Phys. Rev. B \textbf{75},
144420 (2007).

\bibitem{AnRo}F. Anfuso, and A. Rosch, Phys. Rev. B \textbf{76},
085124 (2007).

\bibitem{Mopo}S. Moudgalya, and F. Pollmann, Phys. Rev. B \textbf{91},
155128 (2015).

\bibitem{RSEZ}S.P. Rath, W. Simeth, M. Endres, and W. Zwerger, Ann.
Phys. \textbf{334}, 256-271 (2013).

\bibitem{End}M. Endres, Ph.D. dissertation, 73 (2013).

\bibitem{Nhs}M. Oshikawa, J. Phys.: Condens. Matter \textbf{4}, 7469
(1992); Y. Nishiyama, N. Hatano, and M. Suzuki, J. Phys. Soc. Jpn.
\textbf{65}, 560 (1996).

\bibitem{Schu}H. Schulz, Phys. Rev B \textbf{53}, R2959 (1996). 

\bibitem{BaFi}L. Balents, and M.P. Fisher, Phys. Rev. B \textbf{53},
12133 (1996).

\bibitem{Frad}C. Wu, W.V. Liu, and E. Fradkin, Phys. Rev. B \textbf{68},
115104 (2003).

\bibitem{Furu} M. Tsuchiizu, and A. Furusaki, Phys. Rev. B \textbf{66},
245106 (2002).

\bibitem{BaFi2}H.-H. Lin, L. Balents, and M.P. Fisher, Phys. Rev.
B \textbf{56}, 6569 (1997).

\bibitem{Lede} U. Ledermann, Phys. Rev. B \textbf{64}, 235102 (2001).

\bibitem{Giam} E. Orignac, and T. Giamarchi, Phys. Rev. B \textbf{56},
7167 (1997); see also T. Giamarchi, ``Quantum Physics in One dimension'',
Oxford University Press (2003).

\bibitem{Ners} A.O. Gogolin, A.A. Nersesjan, and A.M. Tsvelik, ``Bosonization
and strongly correlated systems'', Cambridge University Press (1998).

\bibitem{SI} See supplemental material at , where computational details
are provided 

\bibitem{SwSe} B. Swingle, and T. Senthil, Phys. Rev. B \textbf{87},
045123 (2013).

\bibitem{GPMC} D. Greif, M. F. Parsons, A. Mazurenko, C.S. Chiu,
``Site-resolved imaging of a fermionic Mott insulator'', preprint
arXiv:1511.06366

\bibitem{FPO}Y. Fuji, F. Pollmann, M. Oshikawa, Phys. Rev. Lett.
114, 177204 (2015)

\end{thebibliography}

\begin{thebibliography}{10}
\bibitem{NWS1996}R.M. Noack, S.R. White, and D.J. Scalapino, Physica
C \textbf{270}, 281 (1996)

\bibitem{WOHB2001}Z. Weihong, J. Oitmaa, C. J. Hamer, and R. J. Bursill,
J. Phys.: Condens. Matter \textbf{13}, 433 (2001)

\bibitem{MR2012}A. Montorsi, and M. Roncaglia, Phys. Rev. Lett. \textbf{109},
236404 (2012)\end{thebibliography}

\section*{Supplementary Information about DMRG calculations}

\begin{figure}
\includegraphics[width=10cm]{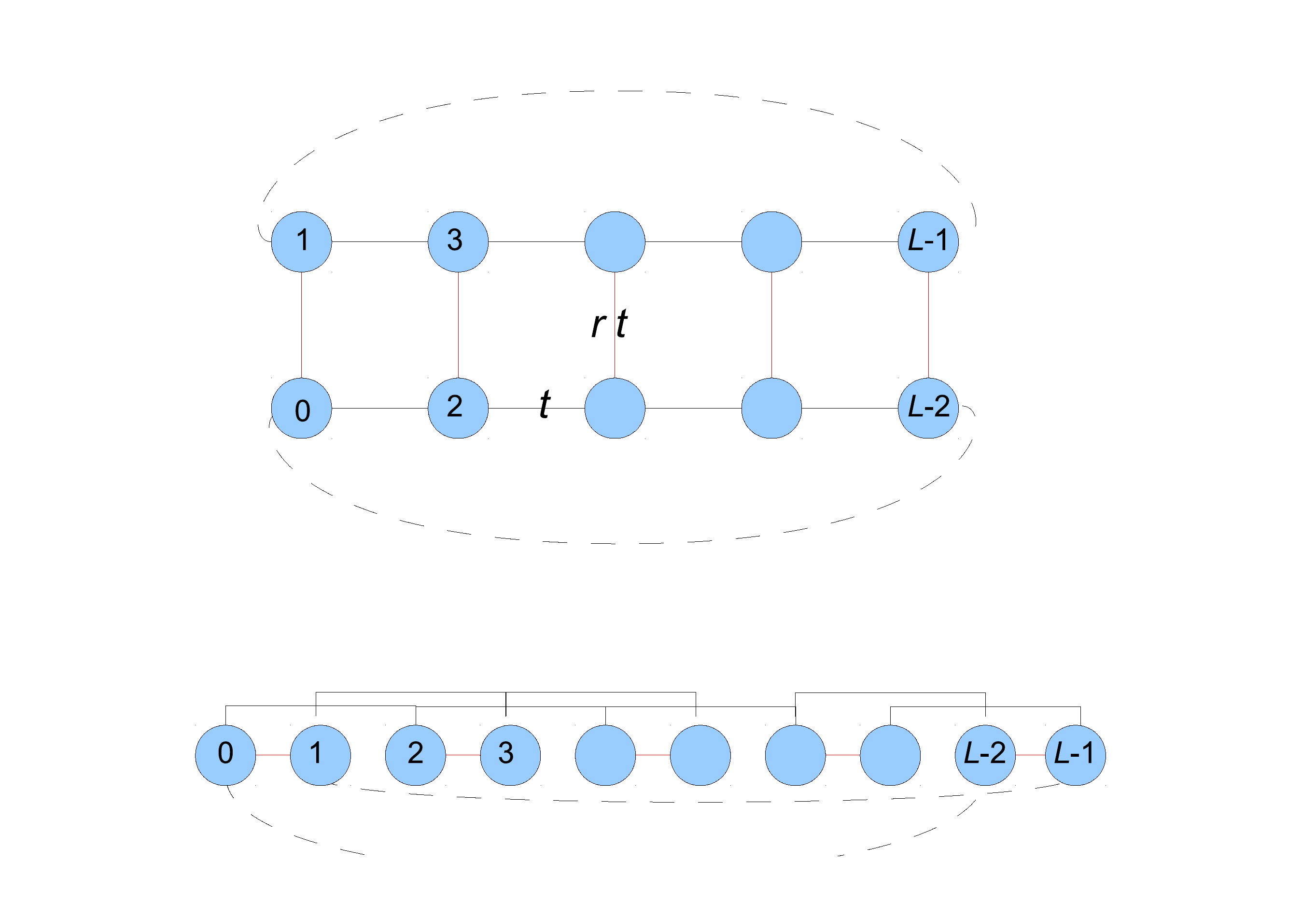}

\caption{Upper panel: Hubbard ladder with two legs $M=2$; Lower panel: Re-destribution
on a 1D chain with nearest (red) and next-to-nearest (black) neighbours
couplings. The dashed lines represent the interactions to be included
in case of PBC. The system has $L=N\times M$ sites in total, that
is $L/2$ rungs. \label{fig:ladderH}}
\end{figure}

In order to handle the problem $M=2$ by means of DMRG we first have
to re-map it into a 1D geometry (see fig. \ref{fig:ladderH}); in
real space this can be done either (a) by tracing a snake-like path
on the ladder or (b) by putting the two sites of a rung inside a virtual
super-site with local dimension of the Hilbert space $D_{{\rm rung}}=16$
and writing all the terms of the Hamiltonian in terms of this new
degrees of freedom. A series of tests however indicated us that option
(b) is not convenient at all from the point of view of the computing
resources to be employed, so here we present only results for option
(a). 

Another issue that is known to be crucial in DMRG calculations are
boundary conditions. While OBC are usually much more favourable because
in a ``block+dot+block+dot'' geometry of the superblock a direct
interaction between blocks is avoided, in this case we have performed
also many calculations with PBC because of two reasons. First, in
the option (a) above the ``price to be paid'' is to have next-to-nearest
neighbours hoppings and so a block-block interaction is unavoidable
in any case. Second, the core of our calculations are parity correlation
functions involving a non-local product of operators between two ``sites''
of the ladder. The drawback of OBC is that there are edge effects
close to the first and last sites, so in order to capture a bulk behaviour
of the correlation functions one has to compute them between sites
that are sufficiently far from the edges. For instance in our case
we have conventionally chosen to compute the correlation functions
from a starting rung at $1/4$ to $3/4$ of the total number of rungs.
In this way we cover half of the ladder length, which is meaningful
distance also with PBC that on the other hand have the advantage of
not involving edge effects. Moreover the analytical results derived
through bosonization formally correspond to PBC geometries.

As regards the number of optimised DMRG states, $m_{{\rm DMRG}}$,
we have performed preliminar tests with up to 2048 states both for
coupled and for uncoupled legs, the latter case being the most severe
one for DMRG because we have essentially two interpenetrating disconnected
chains. If not specified otherwise the results correspond to PBC and
$m_{{\rm DMRG}}=1024$ states that in worst cases (half-filling) yield
a relative energy error of $\sim10^{-3}$ (w.r.t. the exact solution
obtained at $L=16$) after 5 finite-system sweeps.

\subsection*{Parity correlators in ladders ($M=2$)}

\begin{figure}
\includegraphics[width=8cm]{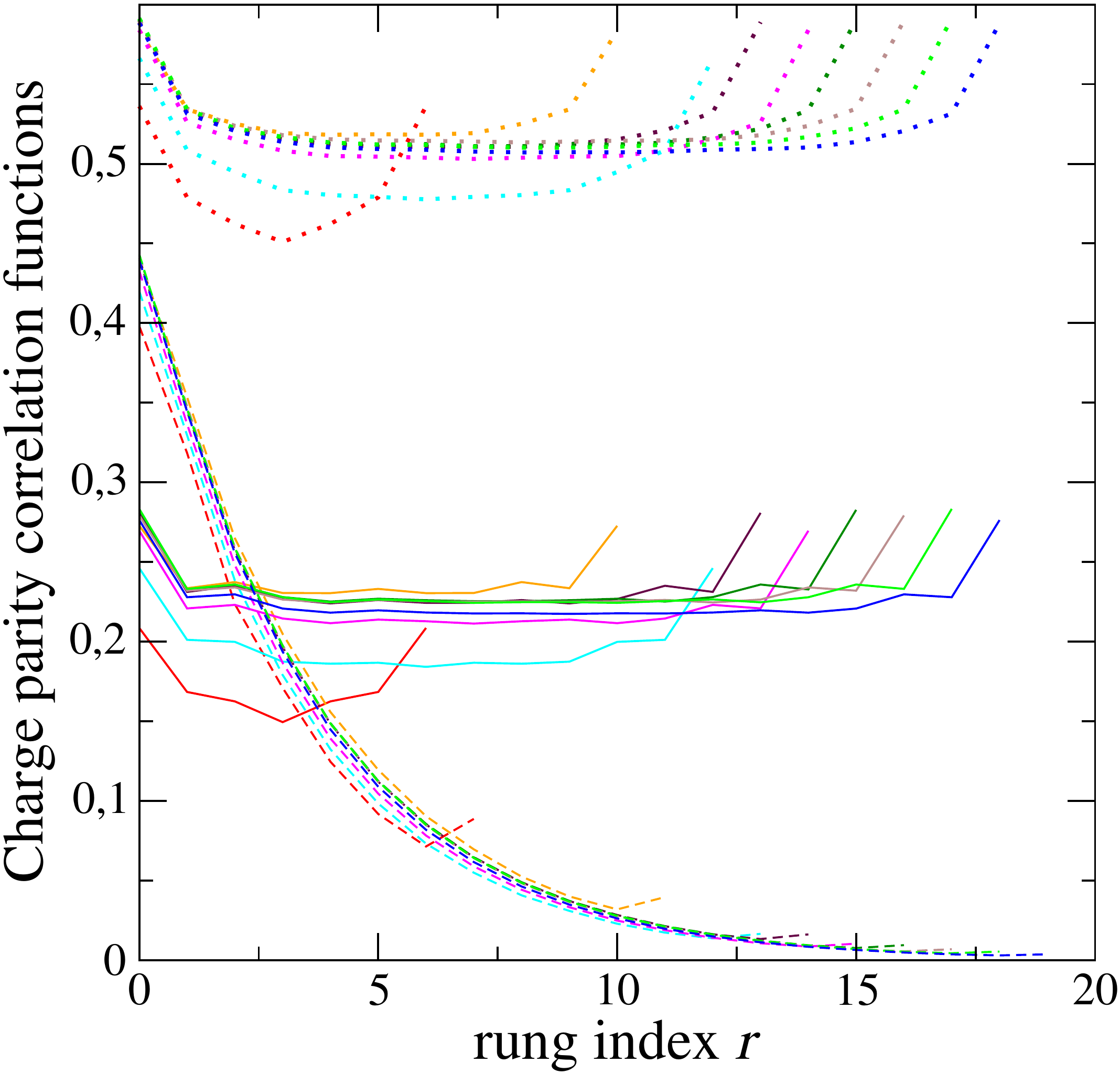}

\caption{Charge parity correlators as functions of the rung distance $r$ for
$U=3$ at half-filling and different systems' sizes ($L=16$ in red,
24 in orange, 28 in cyan, 30 in maroon, 32 in magenta, 34 in dark
green, 36 in brown, 38 in light green, 40 in blue)). Full lines indicate
the non-fractional definition involving all the sites between rung
0 and rung $r$, while the dashed lines involve only the sites of
one leg in the ladder (point-like definition). The dotted line instead
represents the brane fractional definition. \label{fig:chpar_vs_r-U3}}
\end{figure}

As already evident in the 1D case in order to extract the infinite
distance limit of parity correlators one has to investigate first
the dependence on $r$. In the two-legs ladder this dependence turns
out to be nontrivial and also related non-monotonically to the system
size. First we show in fig. \ref{fig:chpar_vs_r-U3} the spatial dependence
of three types of charge parity correlators, namely non-fractional
(solid lines), fractional with factor $1/M=1/2$ in the exponent (dotted
lines), and non-fractional computed on the sites of a leg in the ladder
(dashed lines). While the former two are symmetric with respect to
the central rung in the correlator support, the latter (point-like)
decays to zero for all the values of positive $U$ we have studied.

\begin{figure}
\includegraphics[width=8cm]{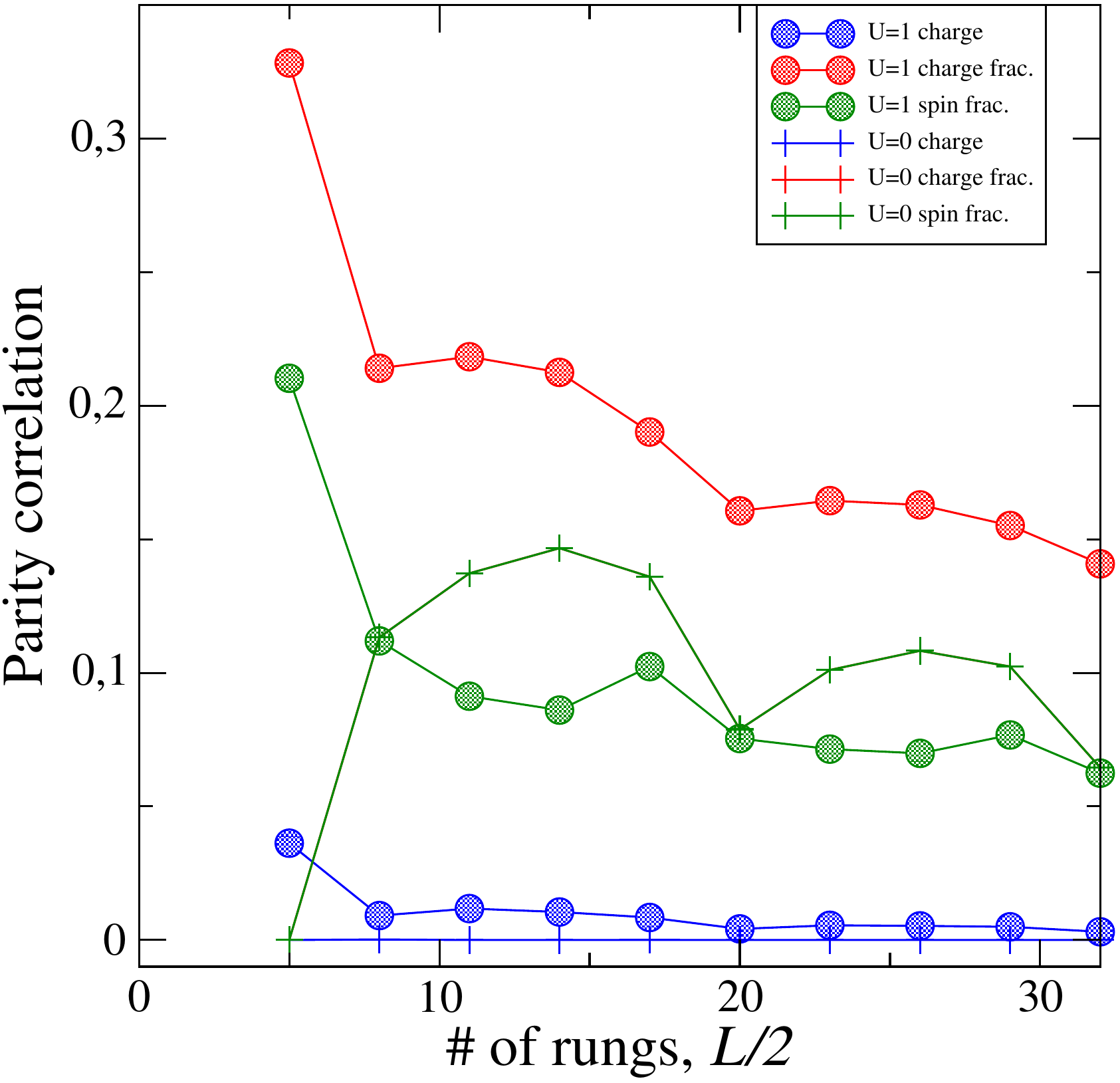}

\caption{Charge parity correlators in the middle of the ladder as functions
of the total rung number $L/2$ for three different choices of $U=1,3,5$
that yield different alternating behaviours (half-filling). Shaded
blue symbols represent non-fractional charge parity while full red
symbols represent the fractional version.\label{fig:parity_vs_size}}
\end{figure}

\begin{figure}
\includegraphics[width=8cm]{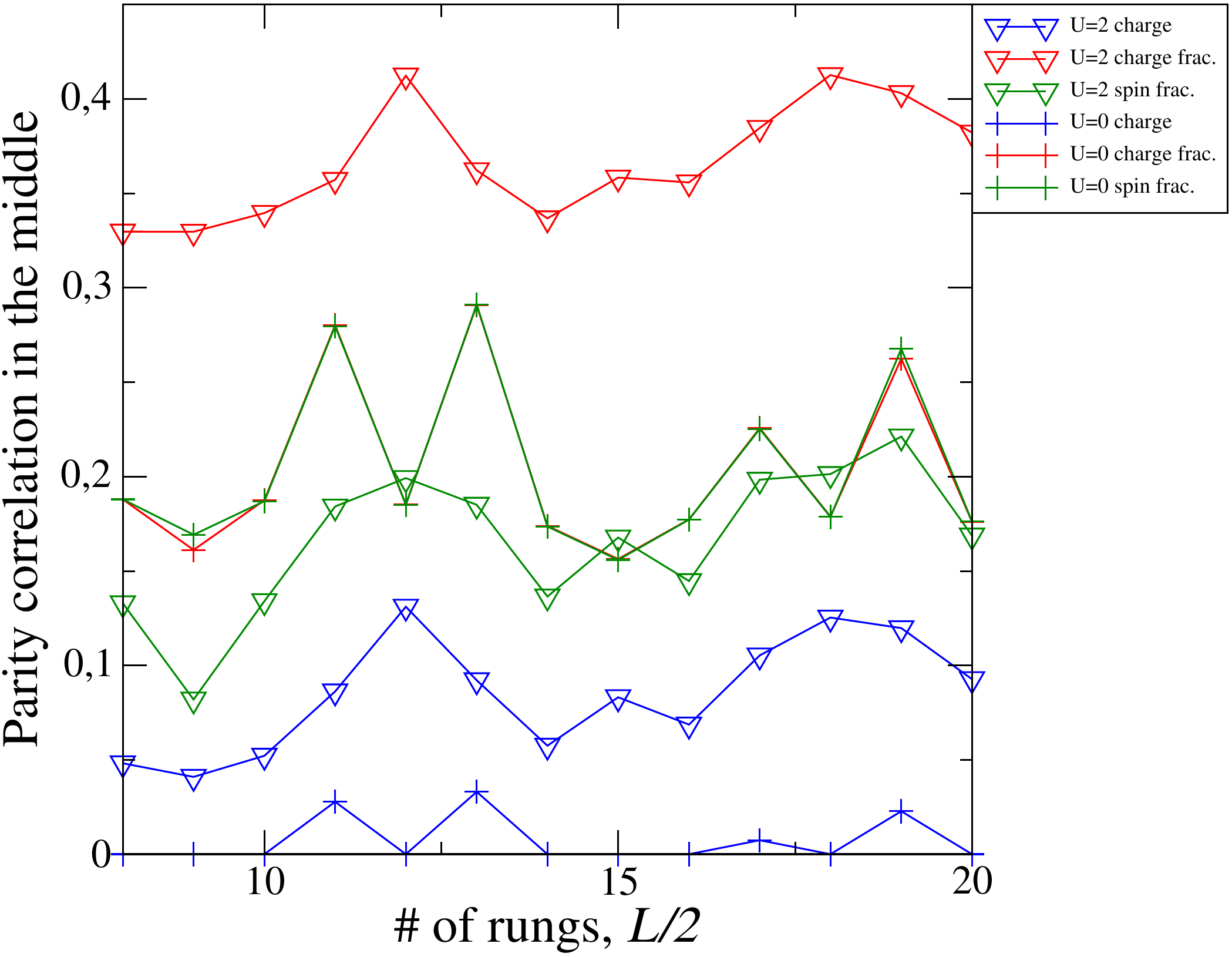}

\caption{Parity correlators in the middle of the ladder as functions of the
total rung number $L/2$ for $U=0$ and $U=2$. The data for the spin
fractional parity are also plotted in green.\label{fig:parity_vs_size02}}
\end{figure}

Moreover, fig. \ref{fig:chpar_vs_r-U3} shows a feature that is present
for all the values of $U$ we have considered: in order to assign
a finite-size value to the parity charge one cannot easily fit an
infinite-distance value from the $r$-dependence because of the non-monotonic
behaviour. Hence, we may conventionally fix the value at the middle
of the curve, at rung number $p$ for ladders of $4p$ or $4p+2$
sites; in the latter case the value in the middle is typically the
same on the central adjacent sites and we have noted that it acquires
a small imaginary part for fractional operator. The representative
central values still depend non-monotonically on the number of sites
$L$. In fig. \ref{fig:parity_vs_size} we present some examples of
this complex behaviour for $U=1,3,5$ to show that the finite-size
sequences seem to have a regularity but with a nontrivial dependence
on the value of $U$. If we may safely conclude that the infinite-length
extrapolations are nonzero for $U\ge3$, the picture again is more
intricate for $U\lesssim2$. The finite-size sequences are plotted
in fig. \ref{fig:parity_vs_size02} and it is hard to decide which
curve will eventually converge to a nonvanishing value for $L\to\infty$.\textbf{\textcolor{red}{{}
}}\textcolor{black}{From our data the point $U=1$ seems to have a
behaviour similar to $U=0$, but with very small or vanishing charge
nonfractional order and with nonzero fractional charge and spin orders.}

\begin{figure}
\includegraphics[width=8cm]{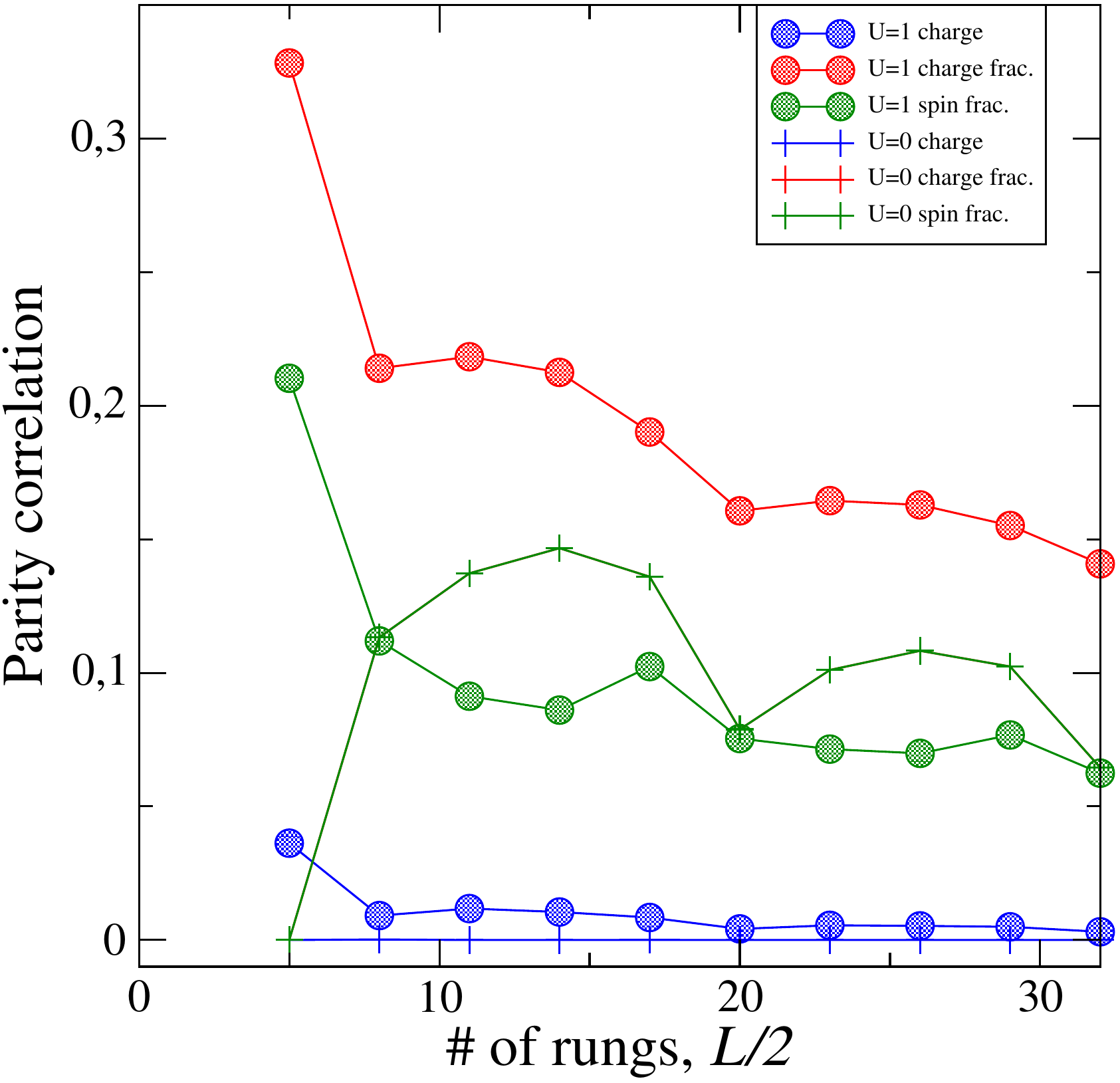}

\caption{Finite-size scaling of the parity correlators from one-quarter and
three-quarters of the ladder length as functions of the total rung
number $L/2$ for $U=0$ and 1 (with OBC).\label{fig:parity-OBC_vs_size}}
\end{figure}

\begin{figure}
\includegraphics[width=8cm]{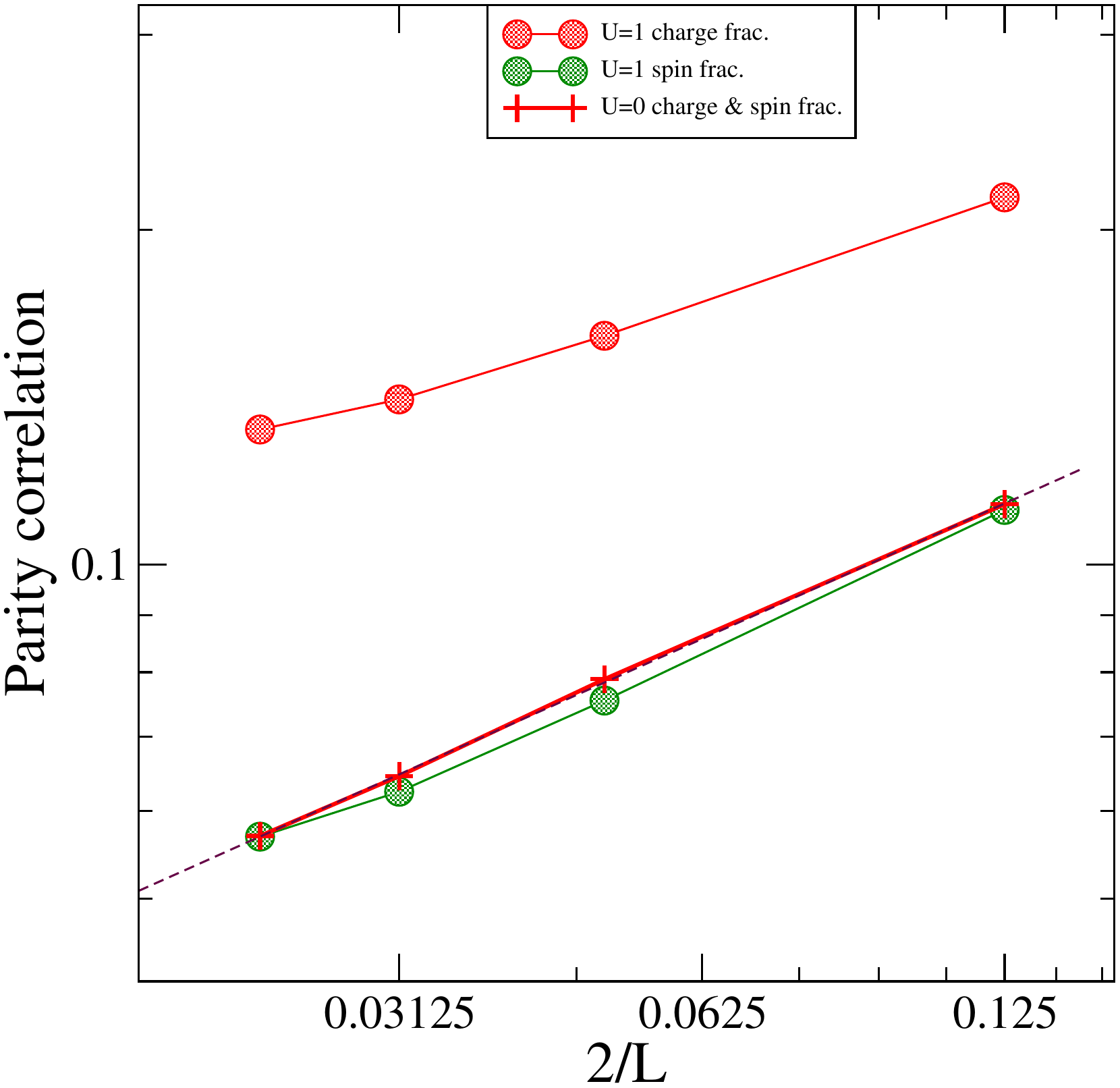}

\caption{As in fig. \ref{fig:parity-OBC_vs_size} in log-log scale, including
$L=88$ with OBC.\label{fig:parity-OBC_vs_invsize}}
\end{figure}

For the reasons above we integrate this series of calculations with
a similar set of data obtained with OBC in order to reach larger sizes;
this part of the study on parity correlators is limited to $U=0$
and $U=1$. The results are displayed in fig. \ref{fig:parity-OBC_vs_size}
for a sequence $L=6p+4$ {[}sizes for which$\Delta_{s}(U=0)=0]$ and
seem to reproduce a super-sequence with period $24$ in $L$; while
the values for the nonfractional parity are compatible with a vanishing
value, the ones for fractional orders may eventually tend to zero
but with a very slow pace. By plotting the local minima (including
the data computed at $L=88$) in log-log scale it seems (fig. \ref{fig:parity-OBC_vs_invsize})
that the curves for $U=1$ show a small upward curvature, while the
data for $U=0$ (coinciding for charge \& spin) scale to zero as $L^{-0.405}$.

\begin{figure}
\includegraphics[width=8cm]{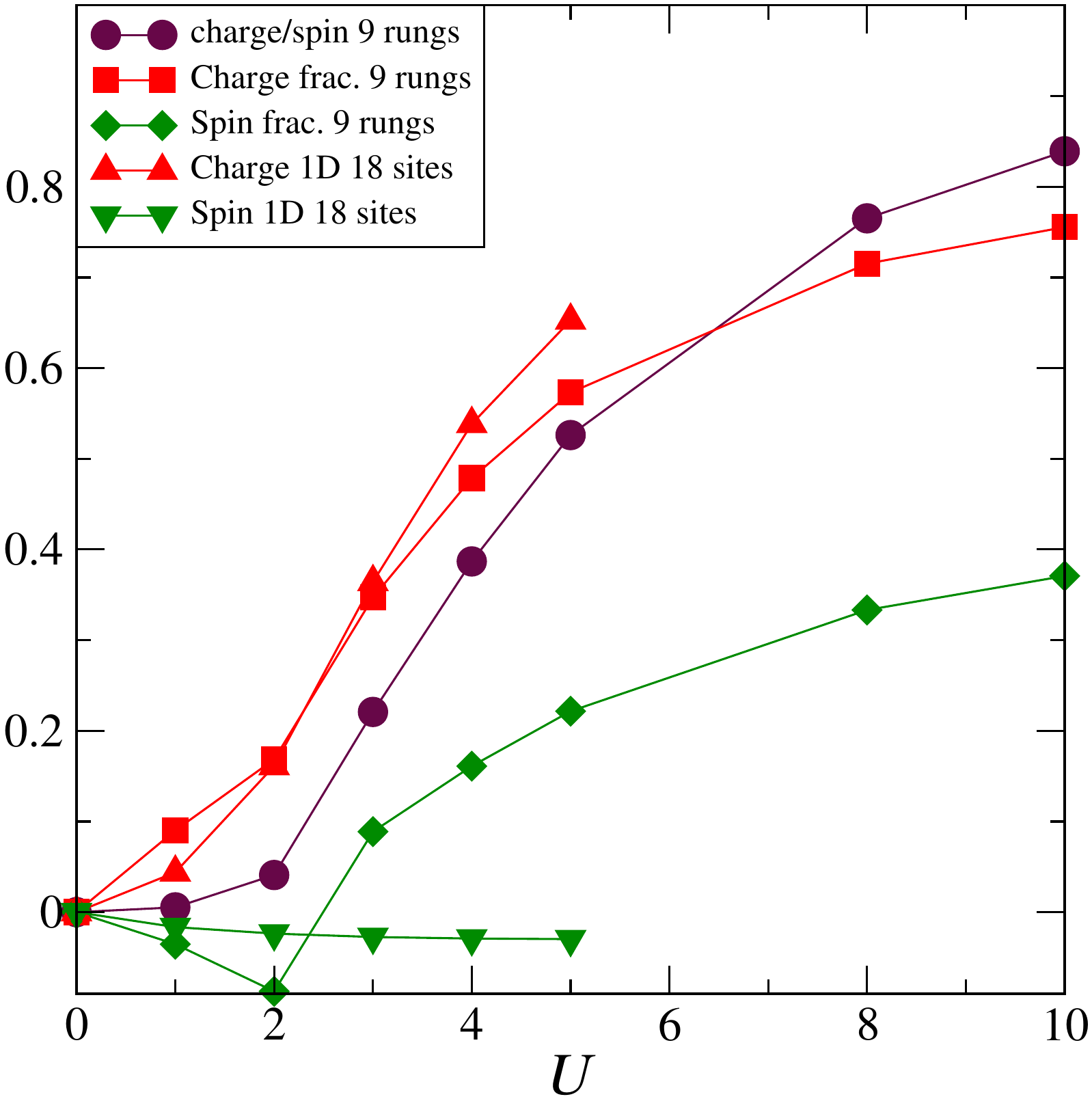}

\caption{Parity correlators (charge and spin, fractional and non-fractional)
in the middle of the ladder as functions of $U$ for $L=18$ and $m_{{\rm DMRG}}=2048$.
Note that the non-fractional spin and charge cases on the ladder (maroon
circles) coincide since $\exp({\rm i}\pi n_{j\downarrow})=\exp(-{\rm i}\pi n_{j\downarrow})$.
The triangles indicate the data for a single Hubbard chain of $18$
sites (see also ref. \cite{MR2012}). (Finite-size finite-values at
$U=0$ have been subtracted out from the corresponding fractional
data, 0.03053 for $M=1$, 0.161 and 0.169 for $M=2$ in charge and
spin resp.)\label{fig:parity_vs_U}}
\end{figure}

Finally, in order to analyse the dependence on $U$ we consider the
various parity correlators taking the rung index $r$ in the middle
of the ladder and then estimating the uniform and oscillating parts
of the correlators through sum and difference of the values at $r$
and $r-1$. This is important because (i) we know that in the 1D case
the behaviour of the spin parity for $U>0$ is purely oscillating
leading to an average vanishing asymptotic order (ii) the decay of
the parity correlators for $M=2$ to their asymptotic values is typically
non-monotonic and in such a way we try to analyse the uniform parts
in a finite-size sample at maximum distance. The results are summarized
in fig. \ref{fig:parity_vs_U} for a choice of large $m_{{\rm DMRG}}=2048$.
In the depicted behaviour there may be important size effects, so
we have performed some additional calculations for larger system's
size $L=38$ (not plotted) and the overall trend is confirmed, even
quantitatively, possibly with the exception of the region $U\lesssim2$
(especially for the fractional spin parity). Already in ref. \cite{NWS1996}
it was pointed out that the DMRG calculations may not be reliable
for such small values of $U$, because the convergence to the ground
state may be poor if the number of sites is too large for the fixed
threshold $m_{{\rm DRMG}}$. 

\begin{figure}
\includegraphics[width=8cm]{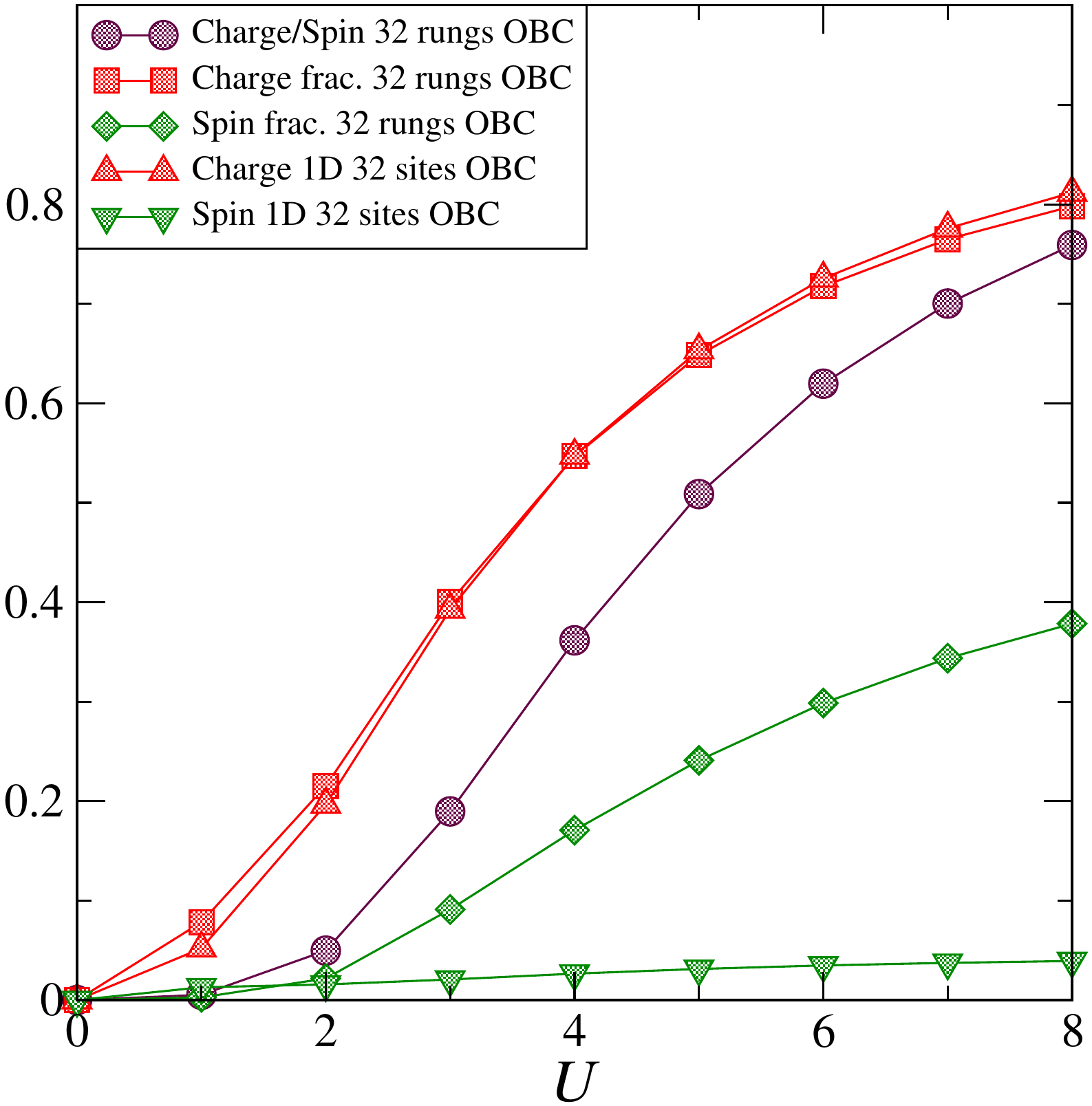}

\caption{Parity correlators (charge and spin, fractional and non-fractional)
from one-quarter and three-quarters of the ladder length $L/2=32$
(with OBC). For comparison the triangles represent the 1D parities
in a chain of 32 sites. (Finite-size finite-values at $U=0$, that
is 0.04695 for $M=1$ and 0.074892 for $M=2$ frac. have been subtracted
out from the corresponding curves.)\label{fig:parity-OBC_vs_U}}
\end{figure}

Hence, we have performed additional calculations for $U=0$ and $U=1$
with OBC and size $L=64$ that, from fig. \ref{fig:parity-OBC_vs_size},
seems to correspond to a sequence of local minima for the fractional
orders and so it is a reasonable upper bound for the infinite-size
curve. The results are reported in fig. \ref{fig:parity-OBC_vs_U},
where the value at $U=0$ should be probably zero in the limit $L\to\infty$
but (as for the spin gap) we do not have a conclusive answer for $U=1$,
even if the data in fig. \ref{fig:parity-OBC_vs_invsize} show a weak
curvature towards positive values in this case.

Already in the 1D case \cite{MR2012} it is seen that the maximum-distance
finite-size value of parity correlators has a nonmonotonic dependence
on $U$, so it is not easy to depict from finite-size data the behaviour
of the parity orders vs $U$. Here, with the aim of comparing the
various definitions, we have conventionally subtracted out the value
of each curve at $U=0$ where the finite-size effect is expected to
be larger. 

\subsection*{Spin Gap}

\begin{figure}
\includegraphics[width=8cm]{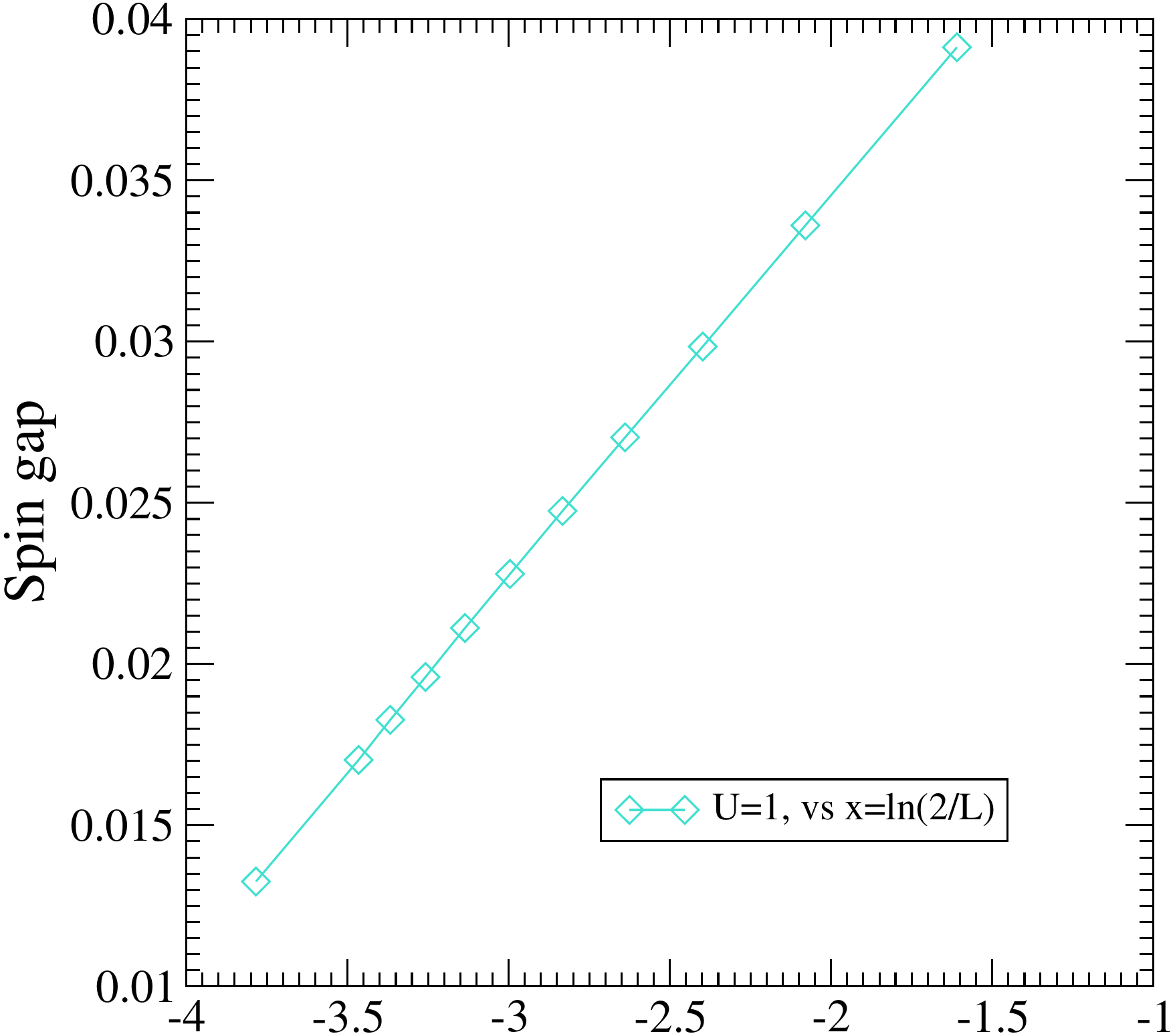}

\caption{Spin gap vs. ladder length for $U=1$ and OBC at half-filling. The
values of $L$ are selected in such a way that the corresponding finite-size
gap at $U=0$ is null (see text). The abscissas in this log scale
suggest that at even larger sizes (left side for which the abscissas
tend to $-\infty$) the data for $U=1$ should eventually ``bend-up''
and settle either to zero or to a small value $\lesssim O(10^{-2})$.
The leftmost point corresponds to $L=88$ (see also above for this
choice).\label{fig:spingap-hfOBCU1}}
\end{figure}

Defined as $\Delta_{s}=E_{L}(N_{{\rm e}},S_{z}=1)-E_{L}(N_{{\rm e}},S_{z}=0)$,
where $E_{L}$ is the lowest energy eigenvalue at a given number of
sites $L$, depending on the particle number $N_{{\rm e}}$ and total
spin $S_{z}$. The study of Weihong and co-workers \cite{WOHB2001}
at half filling puts the boundary of uncertainty about the vanishing
of the spin gap close to $U=1$ (their fig. 6), even if it is also
recalled that bosonization at half-filling predicts a fully gapped
system for $U>0$. In order to inspect the behaviour of the spin gap
at small values of $U$ we concentrate ourselves on the case $U=1$
and compute $\Delta_{s}$ for a sequence of values of $L=6p+4$, $p=1,2,\dots,10$
and OBC (fig. \ref{fig:spingap-hfOBCU1}). This sequence is the one
that gives a vanishing spin gap at finite size for half-filling in
the model. Note that the two non-interacting bands with OBC have the
form $\pm t-2t\cos(k)$ with $k\in(0,\pi)$ in the limit $L\to\infty$,
and at half-filling the ground-state configuration is to fill the
upper band up to $k_{+}=\pi/3$ and the lower band up to $k_{-}=2\pi/3$,
so we can expect to have an oscillatory effect in the results for
$U$ close or exactly zero.

\end{document}